# zIA: a GenAI-powered local auntie assists tourists in Italy

Alexio Cassani[1], Michele Ruberl[1], Antonio Salis[2], Gianluca Boanelli[3] and Giacomo Giannese[3]

[1] Fairmind S.r.l. Società Benefit., via San Vittore 47 – I-20123 Milan, Italy
`{alexio.cassani, michele.ruberl}@fairmind.ai`

[2] Tiscali Italia S.p.A. Loc. Sa Illetta, SS195 km 2,300 – I-09122 Cagliari, Italy
`{antonio.salis}@tiscali.com`

[3] Tiscali Italia S.p.A. V.le Città d'Europa, 681, I-00144 Roma, Italy
`{gianluca.boanelli,giacomo.giannese}@tiscali.com`

**Abstract.** The Tourism and Destination Management Organization (DMO) industry is rapidly evolving to adapt to new technologies and traveler expectations. Generative Artificial Intelligence (AI) offers an astonishing and innovative opportunity to enhance the tourism experience by providing personalized, interactive and engaging assistance. In this article, we propose a generative AI-based chatbot for tourism assistance. The chatbot leverages AI's ability to generate realistic and creative texts, adopting the friendly persona of the well-known Italian all-knowledgeable aunties, to provide tourists with personalized information, tailored and dynamic pre, during and post recommendations and trip plans and personalized itineraries, using both text and voice commands, and supporting different languages to satisfy Italian and foreign tourists' expectations. This work is under development in the Molise CTE research project, funded by the Italian Minister of the Economic Growth (MIMIT), with the aim to leverage the best emerging technologies available, such as Cloud and AI to produce state of the art solutions in the Smart City environment.

## 1. Introduction

Generative Artificial Intelligence (GenAI) is a revolutionary technology poised to reshape numerous industries and significantly impact the global economy. Market forecasts for GenAI indicate exponential growth in coming years: according to Grand View Research estimations [1] GenAI market is expected to reach about 180 billion dollars by 2030, with an annual growth rate of 66%. This rapid expansion is driven by substantial investments from both private and public sectors, fostering the development of innovative technologies and applications. GenAI's potential is particularly evident in media, entertainment and e-commerce, where it can deliver highly personalized content. The abundance of available data serves as a crucial catalyst, enabling the creation and refinement of increasingly sophisticated GenAI models.

According to the UN World Tourism Organization [2] destination management is the coordinated and thoughtful planning of all elements that make up a tourism destination. This can involve anything towards attracting visitors. What a destination management organization (DMO) does is to represent the voice of its destination to potential visitors. It works with travel trade partners to provide travelers with information about the destination before they decide where to go on vacation.

At the same time the concept of Smart Tourism has assumed a crucial relevance for the Smart City. According to the European Commission [3] "a smart city is characterized by a pervasive presence and massive use of information technologies to achieve the optimization of resources, effective and fair governance, sustainability and quality of life, with applications in various fields such as mobility, life, people, governance, economy and environment. The idea of smart tourism is also the fruit of this concept of smart city, and a smart tourism destination can be defined «as a destination where the various stakeholders (…) facilitate the access to tourism and hospitality products, services, spaces, and experiences through innovative ICT-based solutions, making tourism sustainable and accessible, and fully exploiting their cultural and creative heritage." This also means «addressing and improving the quality of life of the local population, as it would benefit from sustainable socio-economic development and would be actively involved in the promotion of the digital culture of the area».

Recommender systems have been one research area that supported the development of Smart Tourism. First approaches have been based on Collaborative Filtering [4,5,6,7] and Content based techniques [8,9]. While they have been effective in predicting and suggesting new destinations or activities that tourists have not thought of, thus broadening their travel horizon, they were a bit generic, not enabling strong personalization of customer experience, and not catching nuances of natural language.

Machine Learning based proposals [10,11,12] offered a better experience identifying patterns and trends in tourism data, such as travel preferences and seasonality. Despite these improvements, these approaches were not able to adapt to the individual and dynamic needs of tourists.

A further step has been achieved with the adoption of Neural Network based models [13,14,15]: learning from complex data, modeling non-linear relationships and deep personalization with continuous learning and adaptation are the major improvements, and result in a more engaging and rewarding user experience. At the same time neural networks can be complex to implement and require technical expertise, significant computing resources, and significant amounts of data. Moreover, understanding how AI models reach their conclusions can be difficult, cited as AI Explainability [16], sensitivity to noise and bias in data [17] and ethical issues related to privacy and misuse of personal data [18].

GenAI promises to inherit the benefits of Neural Network approaches, while improving on the previous issues. Some works and research have been done mostly with the use of ChatGPT [19]. This work is based on a recent experience of part of the authors with the Alpi-GPT implementation [20], a tool developed with OpenAI GPT 3.5 to support travel editors and internal operators of Alpitour, a leading company in the travel sector. While AlpiGPT was specifically designed for "travel editors" – the company experts handling the most intricate customer inquiries – this project focuses on typical travelers visiting Campobasso and Molise region. Given the primarily Italian tourist demographic, with notable international presence, the system will offer both Italian and English text and voice capabilities. It will leverage Large Language Model (LLM)'s general knowledge and comprehensive data on hotels, restaurants, museums, and attractions to satisfy the average visitor's needs.

This paper is structured as follows. Section 2 introduces the solution of the chatbot system, describing the specific requirement and assumptions, driven by the research project, and the unique proposition of the use case. Section 3 provides a detailed description of the system architecture, the use case design, implementation and early tests, including information of the supporting cloud infrastructure. Section 4 gives a description of next steps in terms of work in progress that will be completed by the end of the project. Finally, Section 5 describes the benefits of the developed use case, the exploitation opportunities, and concludes the paper.

## 2. The solution

The identified solution consists of an advanced Chatbot, i.e. a software that simulates and processes human conversations, written and/or spoken, allowing travelers to interact with digital devices as if they were communicating with a real person. They are therefore digital assistants that learn and evolve to provide increasing levels of customization when collecting and processing information. The chatbot has been implemented to manage the engagement funnel, i.e. an engagement model to build a relationship with users-visitors in the province of Campobasso, and with the aim of accompanying them in the search for specific itineraries offered, in planning their trip, for the customization of their travel experience, all using a conversational interface in self-service mode. The perceived friendliness is a design cornerstone, declined in both in the UI graphical style and in the generated text's tone of voice, which mimics the one of an Italian aunt (zIA was chosen as a name because it contains the initials for Artificial Intelligence in Italian, also translates as "auntie").

So the objective is to build a GenAI application, with an advanced App on Smartphone, that integrates one or more language models and through fine-tuning, prompt engineering and RAG techniques will be aligned with the following business requirements:

- **Travel Research and Planning**:
    - Artificial Intelligence (AI) assists in planning vacations, suggesting destinations based on personal preferences and interests,
    - Interaction with the AI is possible through text or voice inputs;
    - The user specifies the type of vacation desired, activities of interest and available budget,
    - The AI provides detailed suggestions regarding places to visit, activities to do and accommodation options.

- **Personalized Itineraries**:
  - After choosing the destination, the AI elaborates personalized itineraries considering the length of stay, personal preferences and local attractions,
  - The AI proposes daily activities, optimal times to visit attractions, recommended restaurants;
  - The itinerary includes personalized suggestions, improving the travel experience.
- Search for tourist itineraries based on criteria such as location, theme (cultural, food and wine, nature, etc.), duration and difficulty level;
- Plan the trip in its fundamental steps such as, for example, defining the destination, choosing the dates, identifying the main attractions, booking flights and accommodations, and finally organizing local transport and meals);
- collect personal preferences, suggest suitable destinations, offer personalized accommodation and transport options, propose unique activities, and provide real-time assistance and advice during the trip.

To effectively implement the solution, a comprehensive assessment of available and official knowledge sources, including both static datasets and dynamic information, is necessary to determine feasibility.

**Data Sources**

The performance of the chatbot strongly depends on the quantity and quality of the data on which to perform training and fine tuning. The following data sources have been identified:
- Documents provided by fellows at the University of Molise (UniMol): these documents have been organized in a directory accessible via FTP or http, and further exposed via web server so that images can be called up from the front-end of the web app. Given the static nature of the data, it is expected that the data ingestion will be launched in a supervised manner, excluding automatic logic, in order to guarantee better quality.
- Third-party sites: the list of websites to crawl has been set temporarily, as suggested by the Municipality of Campobasso, with best websites dedicated to the tourism and cultural promotion of the Molise region. At the moment a one-off crawl is assumed, or in any case to be launched in a supervised manner, excluding automatic logic, as for the previous point.
- Day-by-day events (phase 2): after the first release, a database will be set up to be accessed with a frequency to be defined, assumed daily, for the automated ingestion.
- The www.visitcampobasso.it website: This been identified as the primary container of the data to be acquired, for which access to the Wordpress database has been set up for one-off data ingestion, or in case of difficulty, a crawling will be set up during low-traffic hours, with a possible automated ingestion only in phase 2. Evaluations of alternative knowledge sources are underway, and for each alternative the methods and frequency of acquisition of specific information are being analyzed.
- Google API: it is planned to use calls to the Google Places and Google Maps APIs, both server-to-server as input to the RAG of live data on points of interest for prompt augmentation, and from the client to display maps and data on points of interest.

**User Experience**

After a series of meetings, following requirements expressed by representatives from the Comune of Campobasso, the user-visitor profile (personas) has been defined, very oriented towards the typical tourist from Molise. The following text summarizes the user profile:

*"Antonio improvised a trip from Bologna to Puglia with Giovanna, his Venetian girlfriend, and decided at the last minute to stop in Campobasso, his hometown. His main experience is to create an authentic and meaningful experience for his girlfriend by showing her special places and local traditions. The boy needs immediate and personalized support almost as if it were the support of a relative on site who knows Campobasso well and can offer advice based on his detailed knowledge of the city and the boy's preferences. One of the goals is to optimize the time between family and exploration by creating an unforgettable experience for both."*

**Voice**

The selection of speech-to-text and text-to-speech tools with adequate performance for Italian and English, the languages that the chatbot will support, is still underway. At the moment, the use of the smartphone's native libraries is being excluded. Tone of Voice and Copyright still need to be defined and approved.

**Solution Development steps**

The development of the chatbot follows the main steps:
- Front-End application development
  - User Experience chatbot activation interface and chatbot
  - Front-end development based on NextJS technology
- Backend application development
  - Data import script from content provided by the customer, such as the www.visitcampobasso.it website, pdf files, videos, other web content, etc.
  - Speech-to-text integration
  - Basic user profile management
  - Selection of Large Language Models and Embedding Model (favoring open source solutions as much as possible)
  - Alignment of LLMs to business objectives
    o Fine-Tuning
    o Prompt Engineering
    o Prompt Tuning
  - Integration of MindStream to reduce cases of hallucinations and increase the accuracy of integrated models
  - Provision of an API for integration with the front-end
  - Support for installation in a Dedicated Cloud environment
  - LLM output evaluation.

This implementation is exclusively linked to the development of a chatbot aimed at a single user-visitor profile (personas), very oriented towards the typical tourist from Molise. Any additional profiles will be subject to subsequent integration.

Integrations with event booking systems, hotels, restaurants, etc. are not included in this implementation.

**MindStream**

MindStream is an advanced platform provided by FairMind [27], designed for the orchestration and management of Small Language Models (SLMs).

It offers a comprehensive framework for deploying, coordinating, and optimizing multiple SLMs to work together efficiently.

MindStream's innovative architecture enables the creation of sophisticated AI systems that can potentially match or exceed the performance of larger, proprietary models. By leveraging cutting-edge techniques in AI orchestration, MindStream offers greater flexibility, customization, and control over the AI pipeline.

This platform is particularly valuable for organizations looking to harness the power of AI while maintaining data privacy, ensuring ethical use, and tailoring solutions to specific industry needs.

## 3. Design, Implementation and Tests

The picture shown in figure 1 represents the overall architecture: the creation of an autonomous preliminary web app (with a final target with an Android based App) has been defined, in orange in the diagram below, which maintains the state of the client-side session (cookie) and a representation of the session conversation history on the client and/or server. The installation of a reverse proxy component is also planned to shield access from the public internet to the business logic, on a DMZ machine in purple in the diagram. The installation of the data ingestion and application business logic components (data pipelines, crawler, extractor, third-party API access and API exposure to the client via proxy), in blue in the diagram, and the MindStream product, in black in the diagram, is planned in the back-end machine. A Docker-based container approach will be used to isolate these components. The back-end machine will also be used for testing activities of the infrastructure for fine tuning the models.

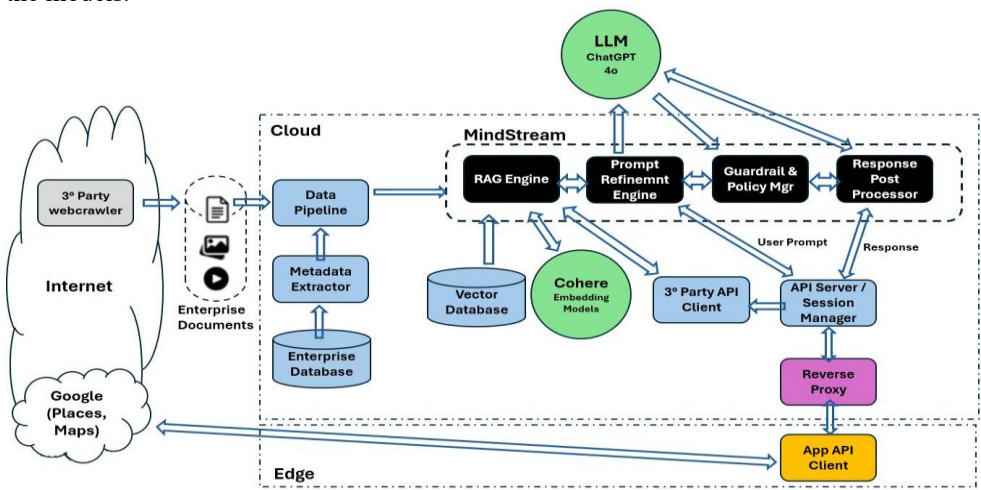

Figure 1 – Chatbot System Architecture

**Cloud Support**

Both front-end and backend machines for the AI chatbot have been installed within the Reevo Cloud platform. Reevo is an Italian cloud provider, with four data centers in Italy, compliant with the highest standards and certifications, focused on infrastructure, storage and cybersecurity services. The IaaS infrastructure guarantees a high level of data protection, with 3 levels of backup: primary storage snapshot on an hourly basis, daily secondary storage backup, and hourly snapshot copy on another datacenter.

The dedicated instance for the system includes two separate Vlans for Fron-end and Back-end, dedicated output bandwidth of 240Mbps, 40 vCPUs at 2.5 Ghz, with 80GB RAM and 1TB of Full Flash storage.

**LLM Models**

Given the continuous evolution of the state of the art regarding LLM models, at the moment the final choice on the model to adopt has not been made; therefore, the first activities aimed at the implementation of the Minimum Viable Product (MVP) have used external calls to OpenAI GPT-4o as the main LLM, and Cohere as the Embedding Model, shown in green in figure1, both as-a-service, with prompt engineering and tuning activities in support. In parallel and separately from the creation of the MVP, fine tuning activities will be carried out on an open-source model, yet to be selected, to test the performance of the infrastructure on the Reevo cloud, and to address any subsequent choices aimed at the creation of a proprietary asset.

**App Development**

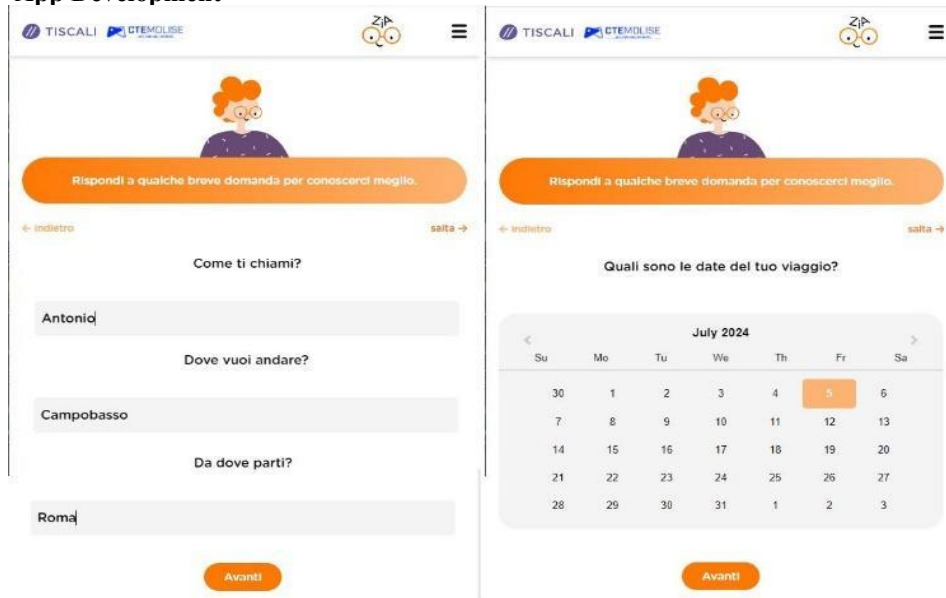

Figure 2 – data collection on destination and dates

The developed chatbot is based on two moments of dialogue with the user: in the first, the AI Assistant acquires basic information about the number of visitors, destination, preferences, and spending limits, as shown in figure 2 and figure 3.

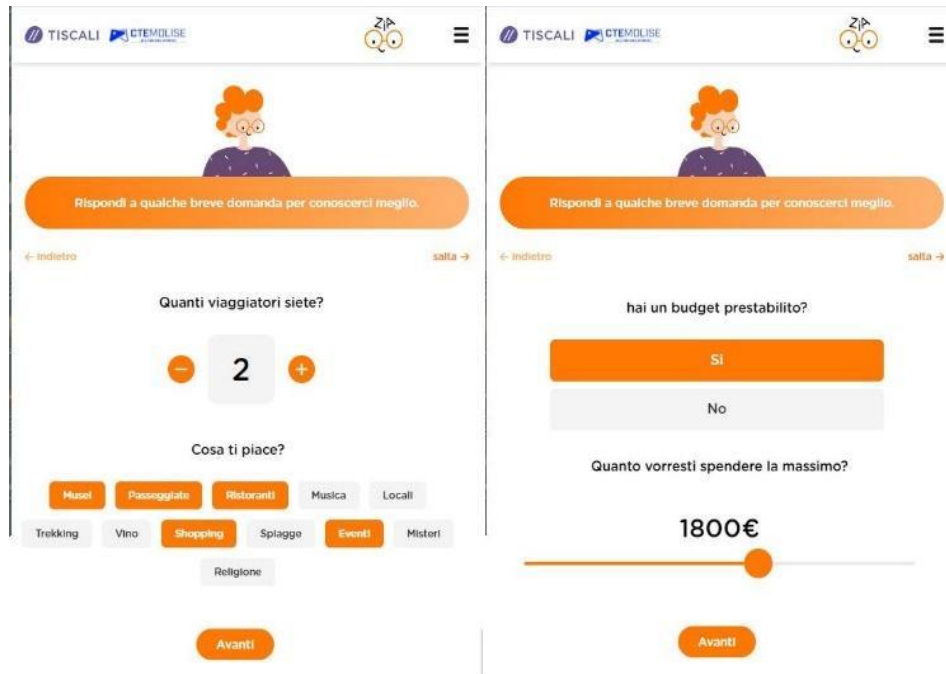

Figure 3 – data collection of number of travelers, preferences and budget

Based on this information, the System builds a knowledge base that guides it in the interactive creation of the holiday with the user. In the second phase, the AI Assistant creates personalized itineraries, taking into account the limits of duration, spending, and all the preferences expressed by the user, arriving at generating optimal routes to optimize time and make the holiday a unique and unforgettable experience.

It's important to highlight that all the interaction is fully dynamic: starting from the answer, the AI Assistant creates a dedicated question structure, therefore generative and not pre-generated a priori for everyone, to collect all the requests, preferences and needs regarding the holiday. This method must be very simple and captivating to generate involvement on the part of the user. Once the first phase of information acquisition is completed, the AI Assistant begins to propose a series of suggestions to the user to prepare the holiday, as shown in figure 4.

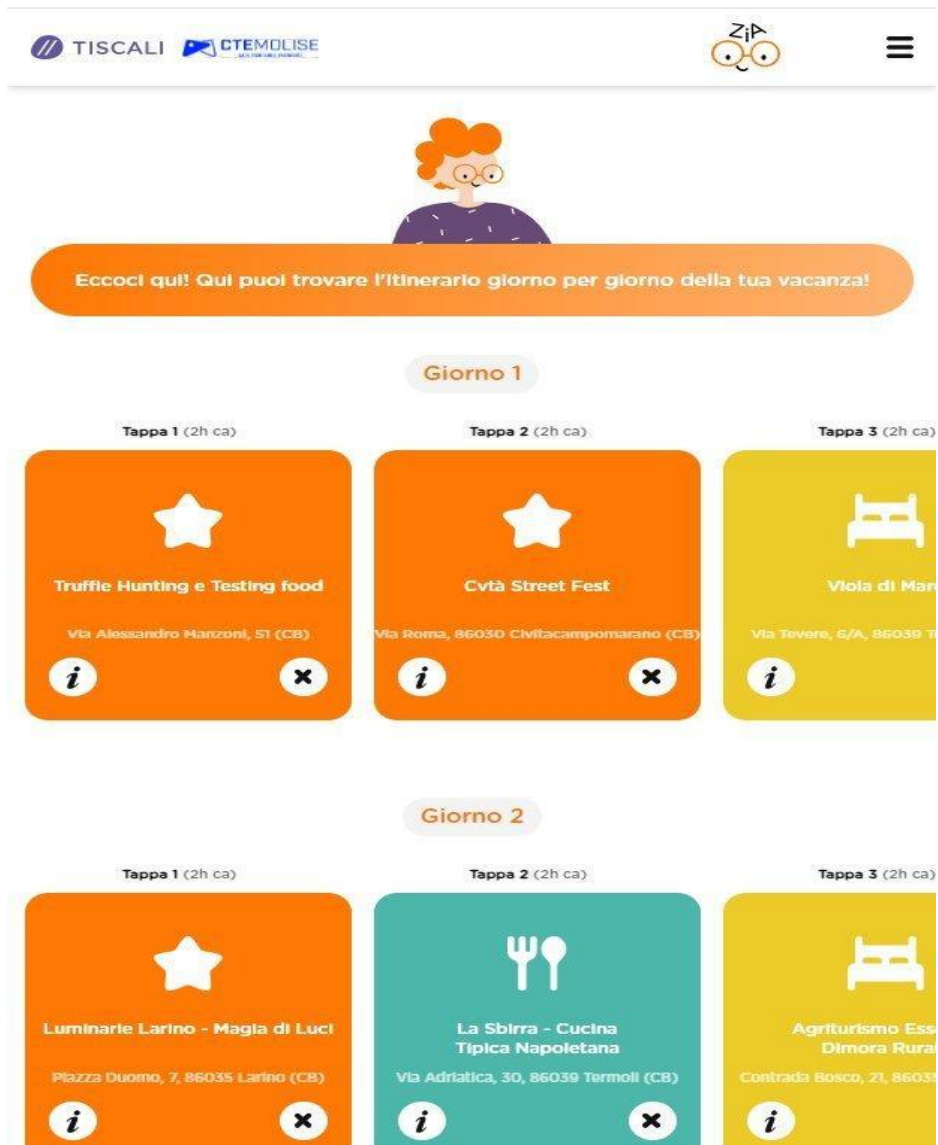
Figure 4 – example of a proposed itinerary

In this phase, the AI will play a key role, building a user profile and comparing it with others on the web, so that it will formulate proposals based on the similarity of tastes and preferences of "similar" users. The user will be able to interact with the AI Assistant, changing or choosing between the options, and the system will keep track of all interactions between the user and the AI Assistant made both when planning the trip and at the end of the trip itself, being able to collect useful feedback to refine the proposals to other users.

The suggestions will include daily activities, visiting times for attractions, museums or particular events, recommended restaurants, taking into account any limitations (allergies, diets, vegetarian regime, etc.), and even optimal routes to save time while traveling.

At the end of the session, the user will be able to decide to download all the information collected through the AI Assistant, through a pdf file, and/or share it.

The first prototype is available online (in Italian) at the following address: https://zia.fairmind.ai. Early tests show good response times: to generate a complex itinerary about 10 seconds are needed, very responsive in all other cases. We plan to add a Nvidia A100 GPU by the end of the project to speedup both training, fine tuning and response times.

## 4. Next Steps

The AI Assistant is now at the stage of being a working prototype. We have a roadmap of following development to extend it to perform better and more accurately generate itinerary on the one hand, and better communicate with the users on the other hand. The first objective will be achieved through benchmarking of commercial LLMs against open source alternatives, and Small Language Models (SLM) to be then orchestrated by our agent-based engine MindStream. The second objective will be achieved migrating to a mobile app, then integrating text-to-speech and AR technology.

### Model enhancements

As we continue to develop and refine our AI Assistant, we are exploring the potential of Small Language Models (SLMs) to enhance performance, sustainability, and user experience. SLMs offer several advantages over their larger counterparts:

- Personalization: SLMs are more easily fine-tuned to specific domains or use cases, allowing for better personalization in the context of travel and tourism.
- Sustainability: With lower computational requirements, SLMs consume less power, making them a more environmentally friendly option. This aligns with the growing demand for sustainable tourism solutions.
- Edge Computing Compatibility: The compact nature of SLMs enables edge deployment, potentially allowing integration directly into the possible mobile app. This can significantly improve response times and reduce latency, enhancing the overall user experience.
- Enhanced Data Protection: By processing data on the device, edge-deployed SLMs can offer improved privacy and data protection, addressing growing concerns about data security in the tourism sector.

Furthermore, we are investigating the potential of combining multiple SLMs using agent-based architectures [24] with MindStream. Recent research suggests that ensembles of specialized SLMs, when orchestrated effectively [25,26], can match or even outperform larger, proprietary models in specific tasks. This approach could provide a more flexible, efficient, and potentially more capable solution for our AI Assistant.

In addition to leveraging existing models, we are evaluating the feasibility of pre-training a custom SLM from scratch, specifically tailored for the travel industry. This model would be trained on a curated dataset of travel-related information, potentially including:
- Destination descriptions and travel guides
- Hotel and restaurant reviews
- Transportation schedules and routes
- Cultural and historical information
- Local customs and etiquette

By creating a travel-specific SLM, we aim to enhance the AI Assistant's understanding of tourism-related queries and improve its ability to generate relevant, accurate, and contextually appropriate responses.

As we progress with these model enhancements, we will continually benchmark performance against our current implementation to ensure that any changes result in tangible improvements to the user experience, response quality, and overall system efficiency.

**UX enhancements**

While the application as it is now is a mobile-first web app, a further version of it would benefit greatly from being re-implemented as a standalone mobile application, possibly re-using some of the code by utilizing available frameworks. A mobile application would enable more enhanced in-destination functionality, essentially enabling hyper contextual, turn-by-turn recommendations by utilizing native device's SDK to determine accurate location and being able to promptly communicate with the traveler through notifications. These two possibilities were initially planned during UX designed but were then scraped off current version, because while they are partly achievable for web apps running on a phone as well, they require user consent checks that are nowadays rejected by a majority of users and are thus not reliable.

By leveraging location and notification, the Concierge AI could then update itinerary and recommendations based on user's actual position tracking, and suggest alternatives based on current location and time. LLMs have demonstrated to generate reliable itineraries "zoomed" at city scale, and through the adoption of SLM the model-based specific itinerary generation at turn-by-turn accuracy could be achievable at an acceptable performance.

Another planned enhancement is the implementation of text-to-speech libraries to augment the perceived friendliness of the app, in an only-too-natural evolution of the personification of our auntie's peculiar character. Testing conducted so far on available library point to ElevenLabs AI Voice Generator as the best performing for an Italian nuanced voice, even though we are closely looking at newcomers such as the new Mars 5 [23] which still sound a bit too mechanic in Italian. While concentrating on Italian rendering, both mentioned frameworks are perfectly fine for standard English to cater for the app's audience of international tourists. And as a side note, the fact that LLMs understand input independently from language used and are able to feed text to the user in their language is a crucial factor to the adoption of GenAI technology in contexts like tourism.

Lastly, we plan in future evolutions of the app to explore the advanced UX capabilities of available Advanced Reality SDKs to blend navigation and multimedia content and enable enhanced user interaction.

## 5. Conclusions

GenAI chatbots stand at the forefront of technological advancement in the chatbot space, boasting several advantages over conventional approaches:

**Deep Personalization**: the chatbot can generate personalized and unique responses based on the individual tourist's questions, requests, profile and individual preferences, going beyond simple generic suggestions;

**Natural, engaging and interactive conversation**: the chatbot can interact with tourists in a natural and intuitive way, using natural language and understanding the nuances of human language. This allows you to create a more engaging and pleasant interaction experience;

**Real-time and immediate support**: The chatbot can provide real-time and assistance to tourists, 24/7, instantly responding to their questions and requests, ensuring immediate and continuous support;

**Continuous Learning and constant updates**: the chatbot can be constantly updated with new information, tourism data and features, ensuring an ever-improving experience that keeps up with the latest industry trends.

In addition to current features, the zIA solution based on GenAI could be further improved:

**Integration with booking systems**: once generated the personalized itinerary, the end user could ask to book hotel, flights, trains, buses, tickets for museums, concerts etc. This would improve the customer experience;

**Assistance during travel**: additional assistance during travel could be provided, helping tourists find nearby restaurants, shops, attractions, events and other services, providing directions and translating languages;

**Management of complaints and problems**: the chatbot can support the customer in case of complaints and problems that tourists may encounter during their trip, offering solutions and immediate support, improving their satisfaction and loyalty.

Finally the emergence of SLMs [21] empowers multi-agent architectures, where each agent can specialize in specific tasks like itinerary planning, recommendation generation, or booking assistance. These agents could run on smaller devices, even at the edge, thus enabling the edge-cloud continuum [22]. This is expected to improve the reliability, accuracy and trust in the resulting systems, being more effective in the specific local tourism needs.

In summary the GenAI chatbot represents a significant step forward in tourism assistance. Offering a personalized, engaging, real-time experience and with potential for development in multiple directions, the chatbot has the potential to revolutionize the way tourists plan, experience and remember their travel experiences.

**Acknowledgments** This work is supported by the Molise CTE Project, funded by MIMIT (FSC 2014- 2020), grant #D33B22000060001